 \documentclass[final,5p,times,twocolumn]{elsarticle}

\usepackage{graphicx}
\usepackage{subcaption}
\usepackage{multirow}
\usepackage{wrapfig}
\usepackage{array}

\usepackage{svg}

\usepackage{tabularray}
\usepackage{tabularx}
\usepackage{rotating}
\usepackage{makecell}
\usepackage{booktabs}

\usepackage[utf8]{inputenc}
\usepackage{amsmath}
\usepackage{amsthm}
\usepackage{amsfonts}
\usepackage{epsfig}
\usepackage{psfrag}
\usepackage{pstricks}
\usepackage{algorithm}
\usepackage{stfloats}

\usepackage{url}

\usepackage{cuted}

\usepackage{algorithmicx}
\usepackage{algpseudocode}  
\floatname{algorithm}{Algorithm}

\graphicspath{{./Figures/}}

\usepackage{amssymb}
\usepackage{bm}
\usepackage{hyperref}
\hypersetup{
            colorlinks=true,
            linkcolor=blue,
            anchorcolor=blue,
            citecolor=blue
            }

\biboptions{comma,sort&compress}

\usepackage{enumitem}
\setlist{nosep,topsep=-\parskip}

\usepackage{ulem}

\journal{Accepted by ASME IDETC/CIE 2026}

\begin{document}

\captionsetup[figure]{labelfont={},name={Figure}}

\begin{frontmatter}

\title{Cubic Hermite Lattice Structures}
\author{Yaonaiming Zhao}
\author{Yuntao Ma}
\author{Guoyue Luo}
\author{Qiang Zou\corref{cor}}\ead{qiangzou@cad.zju.edu.cn}

\cortext[cor]{Corresponding author.}
\address{State Key Laboratory of CAD$\&$CG, Zhejiang University, Hangzhou, 310027, China}

\begin{abstract}
Lattice structures are of growing importance in additive manufacturing, where complex internal geometries are increasingly required for lightweight, high surface-to-volume ratios, multifunctionality, and other superior mechanical properties. Conventional lattice modeling methods typically represent struts with simple primitives, such as cylinders or cones, limiting geometric diversity and the design space.
Although recent efforts have increased strut-shape complexity to address this issue, they often do so at the expense of computational efficiency and modeling robustness. As a result, achieving both rich geometric expressiveness and efficient computation remains a challenging problem.
In this paper, we present an implicit modeling method that expands the design and optimization space of lattice structures while preserving the modeling robustness and efficiency of implicit representations. In our method, each strut is defined as a convolution surface over a skeletal graph, and its profile shape is controlled by a cubic Hermite curve. By exploiting the polynomial structure of both the convolution kernel and the cubic Hermite curve-controlled profile, we derive analytical expressions for efficient field evaluation, avoiding costly and unstable numerical computation.
Four case studies have been conducted to validate the proposed method in terms of profile shape diversity, graded lattice modeling, as well as slicing robustness and efficiency.
\end{abstract}


\end{frontmatter}


\section{Introduction}
\label{sec:intro}
The rapid development of additive manufacturing (AM) has enabled the fabrication of objects with highly complex internal architectures, among which lattice structures constitute an important class~\cite{2017_AM_Aremu_voxel_lattice, 2017_Daynes_lattice_optimisation}.
Owing to their attractive properties such as lightweight, high strength, and absorbing energy~\cite{2020_Chen_lattice_property}, lattice structures have emerged as promising engineered materials in applications including aerospace~\cite{2012_lattice_structure_application_aerospace} and biomedical engineering~\cite{2011_Melchels_mesh_lattice}.
From a geometric perspective, lattice structures are commonly defined by a set of nodes and a set of struts connecting node pairs.
The struts are typically modeled as straight cylinders or cones with either constant or linearly varying radii.
Although such representation is simple and intuitive, it offers limited shape expressiveness and restricted control over material distribution in 3D space, which further leads to confined optimization in downstream applications such as meshing, analysis, and fabrication~\cite{2025_CAD_Luo_explicit_subdivision_surface}.

To overcome these limitations, recent research has sought to increase the geometric complexity of struts, particularly through the use of higher-order curves for more expressive strut shapes. A representative example is the QUADOR framework proposed by Gupta et al.~\cite{2018_GUPTA_quador_hub}.
In QUADOR, each strut is modeled as a surface of revolution whose generatrix, or profile curve, is a conic section, extending the conventional linear radius control to a quadratic formulation.
However, the increased geometric flexibility comes at the cost of higher computational complexity, lower efficiency, and reduced robustness. For instance, evaluating intersections between cubic spline surfaces may require complicated iterative subdivision procedures.

To address these challenges, we propose the cubic Hermite lattice modeling method that further increases strut-shape expressiveness while introducing manageable computational complexity. Specifically, the lattice geometry is represented as the level set of an implicit convolution field, and the strut profile is controlled by a cubic Hermite curve.
In addition, we derive analytical expressions of the resulting implicit field, enabling robust and efficient evaluation of the lattice structure.

The following sections begin with a review of related work in Sec.~\ref{sec:related_work}. Then the proposed method is provided in detail in Sec.~\ref{sec:methods}. After that, several case studies and significant application examples are provided in Sec.~\ref{sec:results} to demonstrate the effectiveness. Finally, conclusions on the method’s advantages and limitations are given in Sec.~\ref{sec:conclusion}.

\section{Related Work}
\label{sec:related_work}
This section reviews prior work on lattice representation for geometric modeling. Existing methods can be broadly grouped into explicit representations, which construct the boundary of struts and junctions, and implicit representations, which encode lattices as scalar fields or level sets evaluated through sampling. We summarize representative approaches in both categories in Sections~\ref{sec:explicit} and~\ref{sec:implicit}, respectively. For a more detailed and complete introduction to geometric modeling of lattice structures, we recommend reading the paper~\cite{2025_CAD_zou_review_geometric_modeling}.

\subsection{Explicit Representation}
\label{sec:explicit}

Early lattice modeling pipelines typically regard a single strut as a cylinder or cone, and the node as a sphere. Then the final solid is obtained through Boolean unions. Although conceptually simple, this representation accurately represents the lattice shape. However, computing the intersections of adjacent struts may be ill-conditioned or create non-manifold artifacts, which reduce robustness for downstream meshing and analysis operations.
To mitigate these issues and to expand the design space, subsequent work explored more sophisticated explicit pipelines, including mesh-based construction~\cite{2005_idetc_Wang_mesh_lattice, 2011_Melchels_mesh_lattice, 2017_IJCIM_Vongbunyong_mesh_lattice, 2024_Zou_Meta_meshing}, B-rep-based construction~\cite{2016_Vongbunyong_Brep_lattice, 2024_Zhao_TPMS2STEP}, voxelization~\cite{2017_AM_Aremu_voxel_lattice}, subdivision-surface-based construction~\cite{2018_Savio_explicit_subdivision_surface, 2025_CAD_Luo_explicit_subdivision_surface}, and parametric-curve-driven strut generation (e.g., sweeping a profile curve along an axis)~\cite{2018_GUPTA_quador_hub, 2019_GUPTA_quador_exact_representation, 2019_GUPTA_programmed_lattice, 2020_CAD_Cirak_quador_add_fillets}. These explicit approaches integrate well with commercial CAD/CAM software packages~\cite{2019_GUPTA_quador_exact_representation, 2019_GUPTA_programmed_lattice}, but they often incur higher computational cost for repeated geometric queries (e.g., point-membership classification), since such queries require nontrivial intersection tests, acceleration structures, or repeated evaluation of trimmed boundary entities~\cite{2025_JMD_explicit_intersection}.

\subsection{Implicit Representation}
\label{sec:implicit}

Implicit representations encode a lattice as the level set of a scalar field (e.g., signed distance fields~\cite{2019_Leonardi_ntop_implicit_modeling_method, 2025_Zhang_implicit_sdf}, convolution-function-based field~\cite{2021_JCISE_Liu_convolution_surface, 2022_SMI_blending_convolution_surface}, and other fields described by analytical functions~\cite{2011_Pasko_Frep_implicit,2013_CAD_Fryazinov_Frep_implicit, 2019_Tang_IJAMT_implicit_function_related_to_radius, 2021_CAD_qyounhong_Vrep, 2022_JCDE_Letov_Frep_implicit, 2022_JMD_Letov_Frep_implicit,letov2024geometric}). These implicit fields are always discretized by regularly sampling from the continuous or piecewise-continuous implicit field for diverse downstream tasks, such as ray tracing~\cite{2025_Guo_implicit_field_ray_tracing}, offsetting~\cite{2021_Ding_implicit_offset}, and slicing~\cite{2019_CAD_Feng_implicit_slicing}. In these tasks, the implicit field values at the sample points could be efficiently evaluated by GPU-based parallel computing, which is crucial for scaling the applications to large-scale lattice structures. In practical applications, the commercial software nTopology is well-known for its powerful modeling and rendering tools based on implicit representation.

Another advantage of implicit representation is the robustness of lattice modeling. Based on this representation, the local smooth transition operation at strut junctions can be conducted by directly performing calculations (e.g., simple summation, max/min operators, or other analytical blending operators)~\cite{1995_Bloomenthal_implicit_blending, 2009_CGF_Adrien_implicit_blending,  2013_TOG_Olivier_implicit_blending, 2022_SMI_blending_convolution_surface} on the field values at the sampling points, thus avoiding explicit surface intersections and the topological failures that frequently arise in B-rep Boolean operations. 

However, as far as we know, these methods still have limitations. For example, the profile shape of struts is trivially straight lines. Our approach achieves curved profile shapes by integrating a parametric control curve into the analytical expression of the implicit field, expanding the design space for the profile shape of struts. 

\section{Methods}
\label{sec:methods}

\begin{figure*}[th!]\centering
  \centering
  \includegraphics[width=1\textwidth]{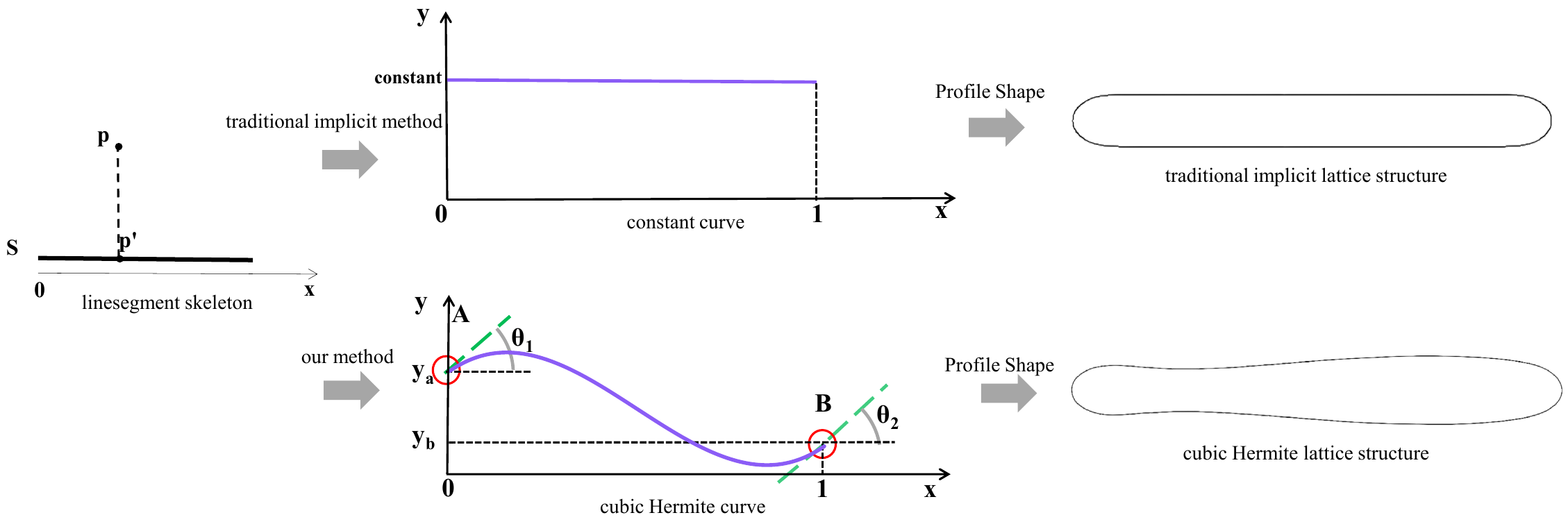}
  \caption{A 1D local coordinate system is constructed on the line segment $\mathbf{S}$ with an axis $\mathbf{x}$. The constant curve represents the straight line profile of the traditional implicit lattice structure (i.e., the distance-field-based lattice structure). The cubic Hermite curve has a start point $\mathbf{A}$ and an end point $\mathbf{B}$, and their coordinates are $(0, y_a)$ and $(1, y_b)$ respectively. $\theta_1$ and $\theta_2$ are the inclination angles of the tangents at points $\mathbf{A}$ and $\mathbf{B}$. The cubic Hermite lattice structure could be derived by replacing the control curve of traditional distance-field-based lattice structures with a cubic Hermite curve. Both curves are reparameterized to the parametric space [0,1] for profile shape control.}
  \label{fig:hermite}
\end{figure*}

Our method decomposes the modeling pipeline into two coupled components.
First, we utilize an implicit representation of lattice structures to avoid the explicit intersection at the junction of struts. Based on this representation, a cubic Hermite curve is adopted to control the profile shape so that each strut can exhibit a non-uniform radius distribution along its axis. This profile shape control approach is inspired by Glassner et al.~\cite{2014_Glassner_Hermite_curve_blending}, which explicitly controls the blending shape of two circles by Hermite curves.
Second, we represent the entire lattice as the accumulation of implicit fields contributed by individual struts, leveraging the superposition property of the convolution surface as used in~\cite{2021_JCISE_Liu_convolution_surface}, and derive the analytical expression of the implicit field of a single strut.

The following subsections are organized as follows. The cubic Hermite lattice structure is introduced in Sec.~\ref{sec:cubithermite}, and the analytical expression of the implicit field is derived in Sec.~\ref{sec:analytical}. Finally, Sec.~\ref{sec:multiplestruts} provides a description of more complex profile shapes of struts derived by splicing multiple struts along the same axis direction.

\subsection{Cubic Hermite lattice structures}
\label{sec:cubithermite}
In our method, the lattice structure is represented as a graph of interconnected skeletons, denoted as $\Omega$. This graph could be expressed as a set of nodes $\upsilon  $ and a set of edges $ \varepsilon $ connecting these nodes. Formally, we have
\begin{equation}
    \Omega=(\upsilon,  \varepsilon)
\end{equation}
If we define $ \mathbf{v}_i \in \upsilon (i = 1,2,3,...,m)$ as the coordinates of the vertices, and $ \mathbf{e}_j \in \varepsilon (i = 1,2,3,...,n)$ as the two end points associated with a group of control parameters $ \kappa $, we have the formula
\begin{equation}
\begin{aligned}
    &\mathbf{v}_i = (x_i,y_i,z_i) \\
    &\mathbf{e}_j = (\mathbf{v}_{start}, \mathbf{v}_{end}, \kappa)
\end{aligned}
\end{equation}
Given the skeleton representation $\Omega$, the implicit lattice structure solid model could be expressed as
\begin{equation}
    \mathbf{V}(\Omega) = \{ \, \mathbf{p}\, | \,F(\mathbf{p})\le constant \,( \mathbf{\forall p} \in R^3) \, \}
\end{equation}
where $F(\cdot)$ is an implicit function.
Considering an element of the skeleton (i.e., an edge with two nodes, denoted as $\mathbf{S}$), the implicit representation of the single strut could be expressed as
\begin{equation}
\label{eq:implicit_function}
    \mathbf{V}(\mathbf{S}) = \{ \, \mathbf{p}\, | \,F_{primitive}(\mathbf{p})\le constant \,( \mathbf{\forall p} \in R^3) \, \}
\end{equation}
where $F_{primitive}(\cdot)$ refers to the implicit function for one strut.
For a trivial strut with the distance field as the implicit function, $F_{primitive}(\cdot)$ could generally be calculated as the minimum distance between the query point $\mathbf{p}$ and the edge $\mathbf{S}$, that is, $ F_{primitive}(\mathbf{p})=min(\mathbf{p, \mathbf{S}})$. In this case, the radius of the strut along the axis is a constant. To achieve a variable radius, we multiply  the implicit function by an extra weight, denoted as $\kappa(\mathbf{x})$. Here $\mathbf{x}$ refers to the length along the strut axis, which is shown in Fig.~\ref{fig:hermite}. With this weight, the value of the implicit function along the axis is scaled.

Inspired by ~\cite{2014_Glassner_Hermite_curve_blending}, a Hermite curve is adopted as the control curve. This control curve is defined in the parameter space $\mathbf{x}\in[0,1]$, denoted as $H(\cdot)$. The function value (i.e., $H(\mathbf{x})$) is used as the weight, expressed as
\begin{equation}
\begin{aligned}
\label{eq:hermite}
    \kappa(\mathbf{x})=H(\mathbf{x})= &(1-3\mathbf{x}^2+2\mathbf{x}^3)y_a\\
    &+(\mathbf{x}-2\mathbf{x}^2+\mathbf{x}^3)tan\theta_1 \\
    &+(3\mathbf{x}^2-2\mathbf{x}^3)y_b\\
    &+(-\mathbf{x}^2+\mathbf{x}^3)tan\theta_2\\
\end{aligned}
\end{equation}
where $\theta_1$, $\theta_2$, $y_a$, and $y_b$ are the geometric parameters shown in Fig.~\ref{fig:hermite}.
Reparameterization is conducted to fit the length of a single strut.
This cubic Hermite curve can apply higher-order shape control than ~\cite{2018_GUPTA_quador_hub}, yielding more complex profile shapes.
With this weight function, the implicit function could be rewritten as
\begin{equation}
\label{eq:sdf}
    F_{primitive}(\mathbf{p})=\kappa(\mathbf{p}\prime )~ min(\mathbf{p, \mathbf{S}})=H(\mathbf{p\prime}) ~ min(\mathbf{p, \mathbf{S}})
\end{equation}
where $\mathbf{p}\prime$ is the projection of the query point $\mathbf{p}$ on the edge $\mathbf{S}$.

It should be noted that, for both the trivial strut with the distance field as the implicit function and our Hermite-curve-controlled strut, the two rounded ends are naturally formed by the level set of the implicit function, without additional control.

\subsection{Analytical convolutional representation}
\label{sec:analytical}

\begin{figure*}[b!]\centering
  \centering
  \includegraphics[width=0.85\textwidth]{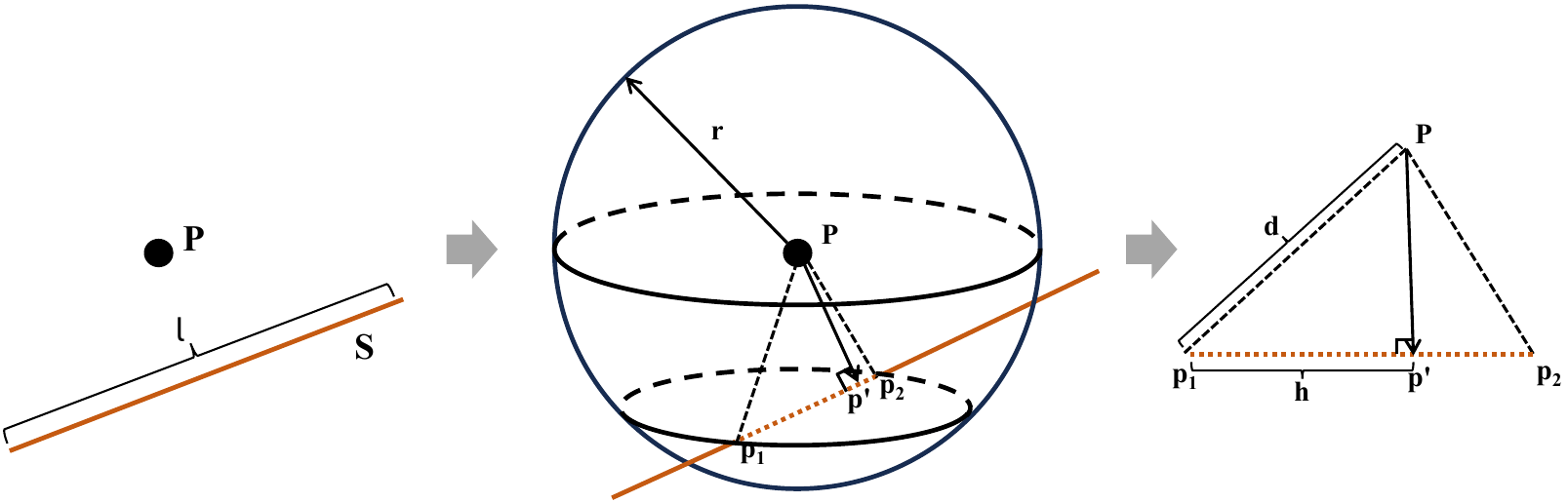}
  \caption{The skeleton of a strut with length $\ell $ (i.e., the line segment $\mathbf{S}$) intersects a ball centered at the query point $\mathbf{P}$ with a radius $\mathbf{r}$ (i.e., the effective radius) at two points $\mathbf{p}_1$ and $\mathbf{p}_2$. $\overrightarrow{\mathbf{P}\mathbf{p\prime}}$ is perpendicular to $\overrightarrow{\mathbf{p}_1\mathbf{p}_2}$.}
  \label{fig:convolution}
\end{figure*}

Since the trivial distance-field-based method has no analytical expression and can only be evaluated on a discrete grid, the computation cost is high. In contrast, we utilize a locally supported convolutional representation~\cite {2021_JCISE_Liu_convolution_surface} to derive an analytical expression for the implicit field, yielding better efficiency and more robust evaluation. To derive the convolution function expression, we consider evaluating the implicit field value of an edge $\mathbf{S}$ at the query point $\mathbf{p}$. As is shown in Fig.~\ref{fig:convolution}, the local support radius is $\mathbf{r}$, and edge $\mathbf{S}$ intersects the ball with radius $\mathbf{r}$ at two points $\mathbf{p}_1$ and $\mathbf{p}_2$.
The convolution function of a single strut is formulated as
\begin{equation}
    F_{primitive}(\mathbf{p})=\int\limits_{\mathbf{S}}^{} f(\mathbf{p}-\mathbf{x} )d\mathbf{x} 
\end{equation}
Here $ f(\mathbf{p}-\mathbf{x})$ refers to the locally supported quartic polynomial kernel function used in ~\cite{2021_JCISE_Liu_convolution_surface}, which is formulated as
\begin{equation}
    f(\mathbf{p}-\mathbf{x})= 
    \begin{cases}
    (1-||\mathbf{p}-\mathbf{x}||^2/\mathbf{r}^2)^2   \quad ||\mathbf{p}-\mathbf{x}||< \mathbf{r}\\
    0 \quad \quad \quad\quad\quad\quad\quad\quad\quad otherwise
\end{cases}
\end{equation}
Substituting the expression of $min(\cdot)$ in Eq.(\ref{eq:sdf}) as this function, Eq.(\ref{eq:sdf}) can be rewritten as
\begin{equation}
    F_{primitive}(\mathbf{p})=\int\limits_{\mathbf{S}}^{} H(\mathbf{x}) f(\mathbf{p}-\mathbf{x} )d\mathbf{x} 
\end{equation}
Since $H(\mathbf{x})$ and $ f(\mathbf{p}-\mathbf{x} )$ are both polynomial functions, the result of the integration (i.e., $F_{primitive}(\mathbf{p})$) is also an analytical polynomial function. The calculation for the integral is as follows
\begin{equation}
\begin{aligned}
    F_{primitive}(\mathbf{p}) &= \int\limits_{l_1}^{l_2}H(\mathbf{x})(1-(\mathbf{p}-\mathbf{x})^2/\mathbf{r}^2)^2dx\\
    &= \frac{1}{\mathbf{r}^4} \int\limits_{l_1}^{l_2}H(\mathbf{x})(\mathbf{r}^2 - (\mathbf{p}-\mathbf{x})^2)^2dx\\
    &= \frac{1}{\mathbf{r}^4} \int\limits_{l_1}^{l_2}H(\mathbf{x})\left \{  \mathbf{r}^2 - \left[ (h-x)^2 + (d^2-h^2) \right]\right \}^2dx \\
    &= \frac{1}{\mathbf{r}^4}\int\limits_{l_1}^{l_2}H(\mathbf{x}) (\mathbf{r}^2 -x^2+2xh-d^2)^2dx \\
\end{aligned}
\end{equation}

Substituting $H(\mathbf{x})$ with the expression in Eq.~\ref{eq:hermite}, we can get the final formula of the implicit field
\begin{equation}
\begin{aligned}
F_{primitive}(\mathbf{p}) &= \frac{1}{\mathbf{r}^4} \int\limits_{l_1}^{l_2} [ (1-3(x/\ell)^2+2(x/\ell)^3)y_a \\
    &+((x/\ell)-2(x/\ell)^2+(x/\ell)^3)tan\theta_1 +(3(x/\ell)^2-2(x/\ell)^3)y_b\\
    &+(-(x/\ell)^2+(x/\ell)^3)tan\theta_2 ] (\mathbf{r}^2 -x^2+2xh-d^2)^2dx \\
    & =\frac{1}{\mathbf{r}^4}  \sum_{i=0}^{3} \int\limits_{l_1}^{l_2} \kappa_ix^i(\mathbf{r}^2 -x^2+2xh-d^2)^2dx \\
    & =\frac{1}{\mathbf{r}^4}  (F_1(\mathbf{p})+F_2(\mathbf{p})+F_3(\mathbf{p})+F_4(\mathbf{p})) \\
    & \kappa_0 = y_a \\
    & \kappa_1 = tan\theta_1/ \ell \\
    & \kappa_2 = (-3y_a-2tan\theta_1+3y_b-tan\theta_2) / \ell^2\\
    & \kappa_3 = (2y_a+tan\theta_1-2y_b+tan\theta_2) / \ell^3
\end{aligned}
\end{equation}
where $\ell$ is the length of the line segment. Using polynomial integration, we can calculate
\begin{equation}
\begin{aligned}
F_1(\mathbf{p}) =&\frac{1}{\mathbf{r}^4} \int\limits_{l_1}^{l_2}\kappa_0(\mathbf{r}^2 - x^2 + 2xh - d^2)^2dx\\
=& \frac{1}{\mathbf{r}^4} \{  (\mathbf{r}^2 - d^2)^2 (l_2 - l_1)\\
&+ 2 h (\mathbf{r}^2 - d^2) (l_2^2 - l_1^2) \\
&+ \frac{2}{3} [2h^2 - (\mathbf{r}^2 - d^2)]  (l_2^3 - l_1^3) \\
& - h(l_2^4 - l_1^4)  +\frac{1}{5} (l_2^5 - l_1^5) \}\\
\end{aligned}
\end{equation}
\begin{equation}
\begin{aligned}
F_2(\mathbf{p}) =&\frac{1}{\mathbf{r}^4} \int\limits_{l_1}^{l_2}\kappa_1x(\mathbf{r}^2 - x^2 + 2xh - d^2)^2dx\\
=&\frac{1}{\mathbf{r}^4} \{\frac{1}{2}  (\mathbf{r}^2 - d^2)^2 (l_2^2 - l_1^2)\\
&+\frac{4}{3} h (\mathbf{r}^2 - d^2) (l_2^3 - l_1^3)\\
&+  [h^2 - \frac{1}{2} (\mathbf{r}^2 - d^2)] (l_2^4 - l_1^4)\\
& - \frac{4}{5} h (l_2^5 - l_1^5)  + \frac{1}{6} (l_2^6 - l_1^6)  \} \\
\end{aligned}
\end{equation}
\begin{equation}
\begin{aligned}
F_3(\mathbf{p}) =&\frac{1}{\mathbf{r}^4} \int\limits_{l_1}^{l_2}\kappa_2x^2(\mathbf{r}^2 - x^2 + 2xh - d^2)^2dx\\
=&\frac{1}{\mathbf{r}^4} \{ \frac{1}{3}  (\mathbf{r}^2 - d^2)^2 (l_2^3 - l_1^3)\\
&+  h (\mathbf{r}^2 - d^2) (l_2^4 - l_1^4)\\
&+ \frac{2}{5}  [2h^2 - (\mathbf{r}^2 - d^2)] (l_2^5 - l_1^5)\\
&- \frac{2}{3}h(l_2^6 - l_1^6)  + \frac{1}{7} (l_2^7 - l_1^7) \}
\end{aligned}
\end{equation}
\begin{equation}
\begin{aligned}
F_4(\mathbf{p}) =&\frac{1}{\mathbf{r}^4} \int\limits_{l_1}^{l_2}\kappa_3x^3(\mathbf{r}^2 - x^2 + 2xh - d^2)^2dx\\
=&\frac{1}{\mathbf{r}^4} \{ \frac{1}{4}  (\mathbf{r}^2 - d^2)^2 (l_2^4 - l_1^4) \\
&+\frac{4}{5} h (\mathbf{r}^2 - d^2) (l_2^5 - l_1^5)\\
&+ \frac{1}{3}  [2h^2 - (\mathbf{r}^2 - d^2)] (l_2^6 - l_1^6)\\
&- \frac{4}{7}h(l_2^7 - l_1^7)  + \frac{1}{8} (l_2^8 - l_1^8)  \}
\end{aligned}
\end{equation}
where $d=||\overrightarrow{\mathbf{p}_1 \mathbf{P}}||$ and $h=||\overrightarrow{\mathbf{p}_1\mathbf{p\prime}}||$, which are shown in Fig.~\ref{fig:convolution}.

\begin{figure}[th!]\centering
  \centering
  \includegraphics[width=0.48\textwidth]{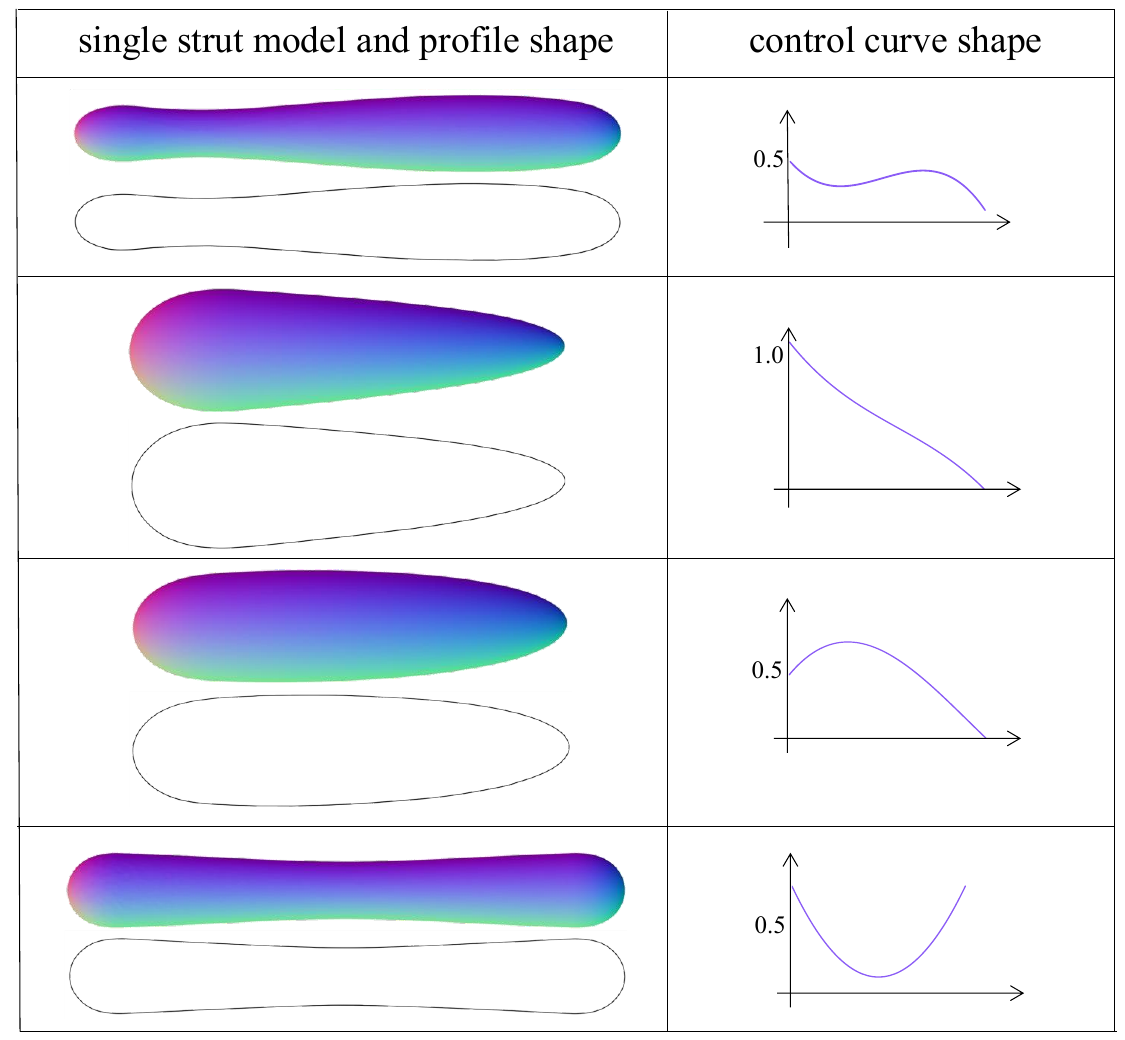}
  \caption{The control curve can only control the field values, not directly control the profile shape.}
  \label{fig:profile_different_control_curve}
\end{figure}

It should be noted that since the shape of the implicit surface is determined jointly by the control curve parameters and the isosurface threshold, the final shape of the implicit surface is not a revolution surface of the Hermite curve, as is shown in Fig.~\ref{fig:profile_different_control_curve}.

\subsection{More complex profile shapes of struts}
\label{sec:multiplestruts}
\begin{figure}[th!]\centering
  \centering
  \includegraphics[width=0.48\textwidth]{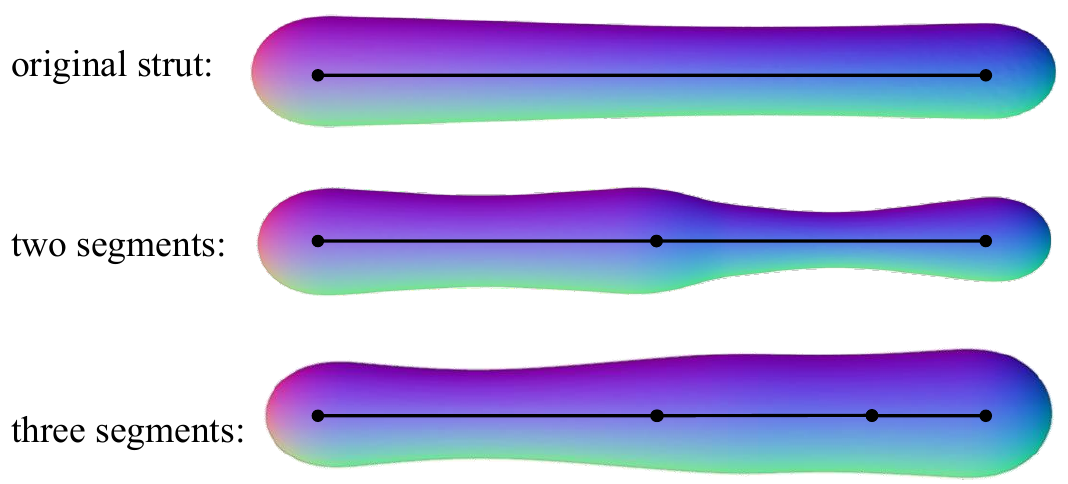}
  \caption{A strut is divided into 2, 3, 4, and 5 segments with independent control parameters.}
  \label{fig:multi_strut}
\end{figure}

In the above subsection, we have achieved the profile shape control of a single strut.
To achieve more complex profile shapes, we divide the line segment of a strut into multiple segments and add individual control parameters to each line segment, as is shown in Fig.~\ref{fig:multi_strut}.

Due to the superposition property of the convolution surface we adopted, the implicit field value could be evaluated by the addition of the field values contributed by each segment. Since the control parameters of different segments are independent, their permutations could greatly enrich the final shape of the strut. Moreover, it should be noted that our method, based on implicit representation, is naturally smooth at the junctions of adjacent struts, without the necessity to consider first-order continuity issues as the biquador method proposed by Gupta et al.~\cite{2018_GUPTA_quador_hub}.

\section{Results and discussions}
\label{sec:results}

\begin{figure}[th!]\centering
  \centering
  \includegraphics[width=0.47\textwidth]{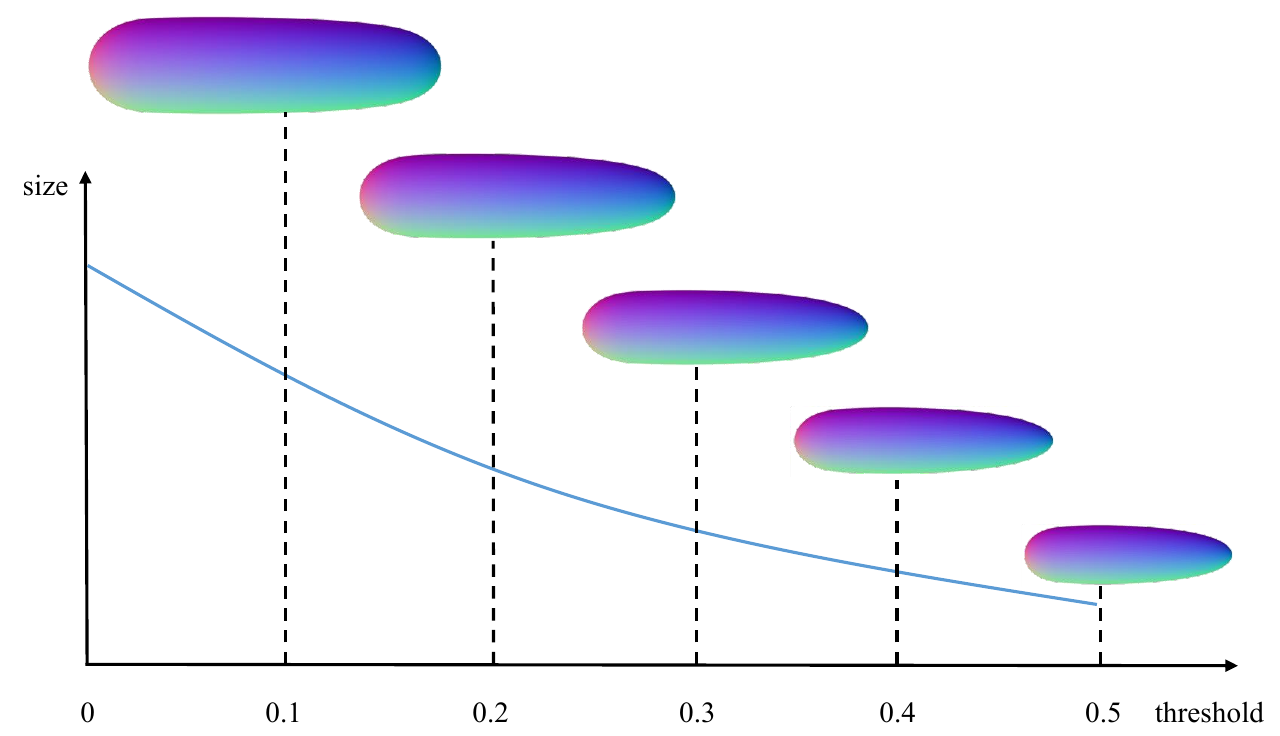}
  \caption{Struts with consecutively varying threshold. The threshold ranges from 0.1 to 0.4, with an interval of 0.1.}
  \label{fig:variable_strut}
\end{figure}

\begin{figure}[th!]\centering
  \centering
  \includegraphics[width=0.45\textwidth]{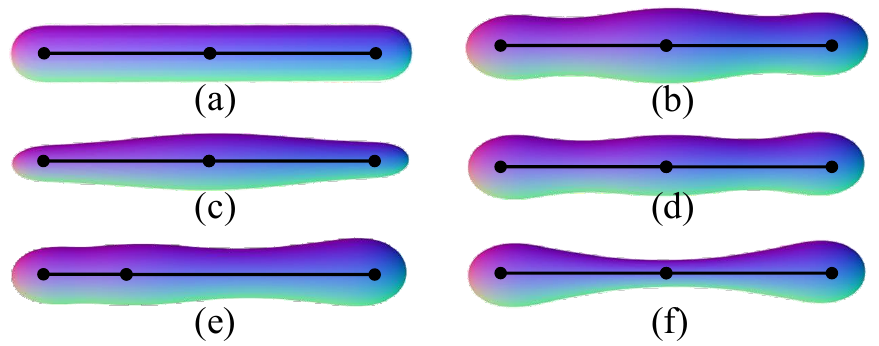}
  \caption{Two struts joined together. (a)-(d) and (f) are constructed by two struts of the same length. (e) consists of two struts with different lengths.}
  \label{fig:double_strut}
\end{figure}

The proposed method has been implemented using C++ on a computer with an Intel Core i9-12900K CPU and 128GB
RAM.  Based on this implementation, two case studies are presented to demonstrate the diverse profile shapes of single struts and the microstructure models constructed by these struts. Moreover, three important applications of our method (graded lattice structure, field-based lattice structure density optimization, and slicing) are presented to demonstrate its potential.
All the implicit fields are discretized and converted into a triangular mesh for presentation using the Marching Cubes algorithm~\cite{1993_Nagae_marching_cubes} with a GPU-based implementation. The slicing algorithm used here is an implicit-representation-based method~\cite{2021_JCISE_Liu_convolution_surface}. For each slicing layer, the values at a grid of sample points are evaluated, and the Marching Square algorithm is utilized to extract the contour. An NVIDIA RTX 3090 GPU is used for parallel computation.

\begin{figure*}[th!]\centering
  \centering
  \includegraphics[width=0.95\textwidth]{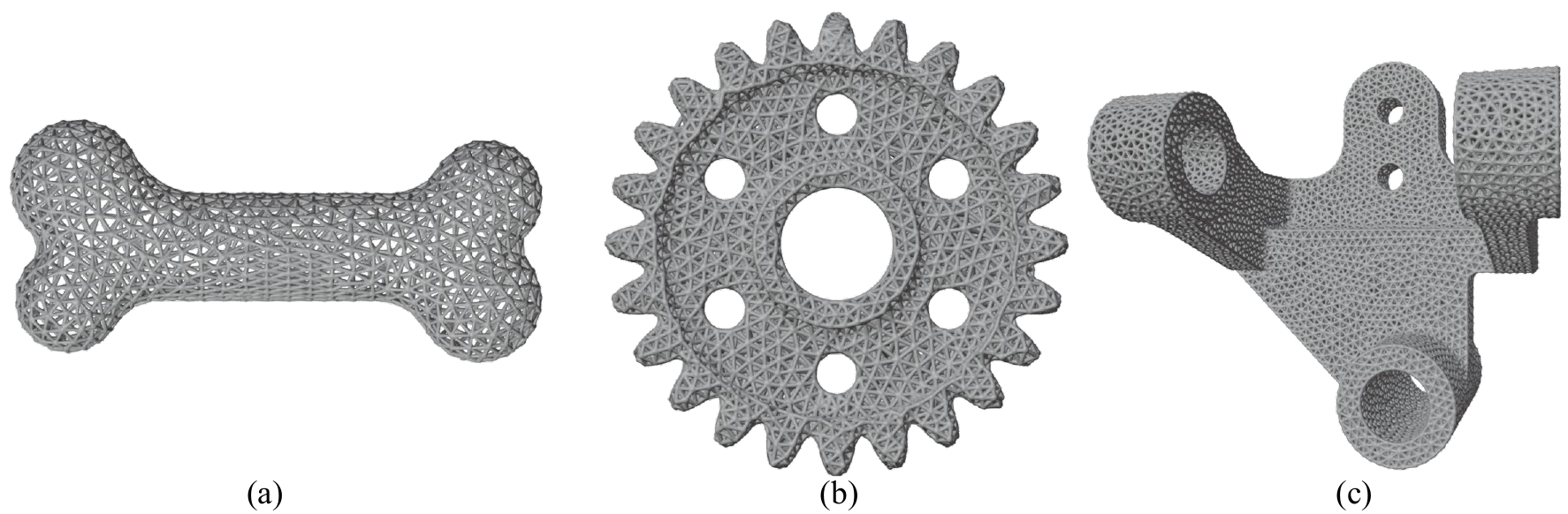}
  \caption{Three cubic Hermite lattice structure models.}
  \label{fig:solid_models}
\end{figure*}

\subsection{Examples}

\begin{figure*}[th!]\centering
  \centering
  \includegraphics[width=0.95\textwidth]{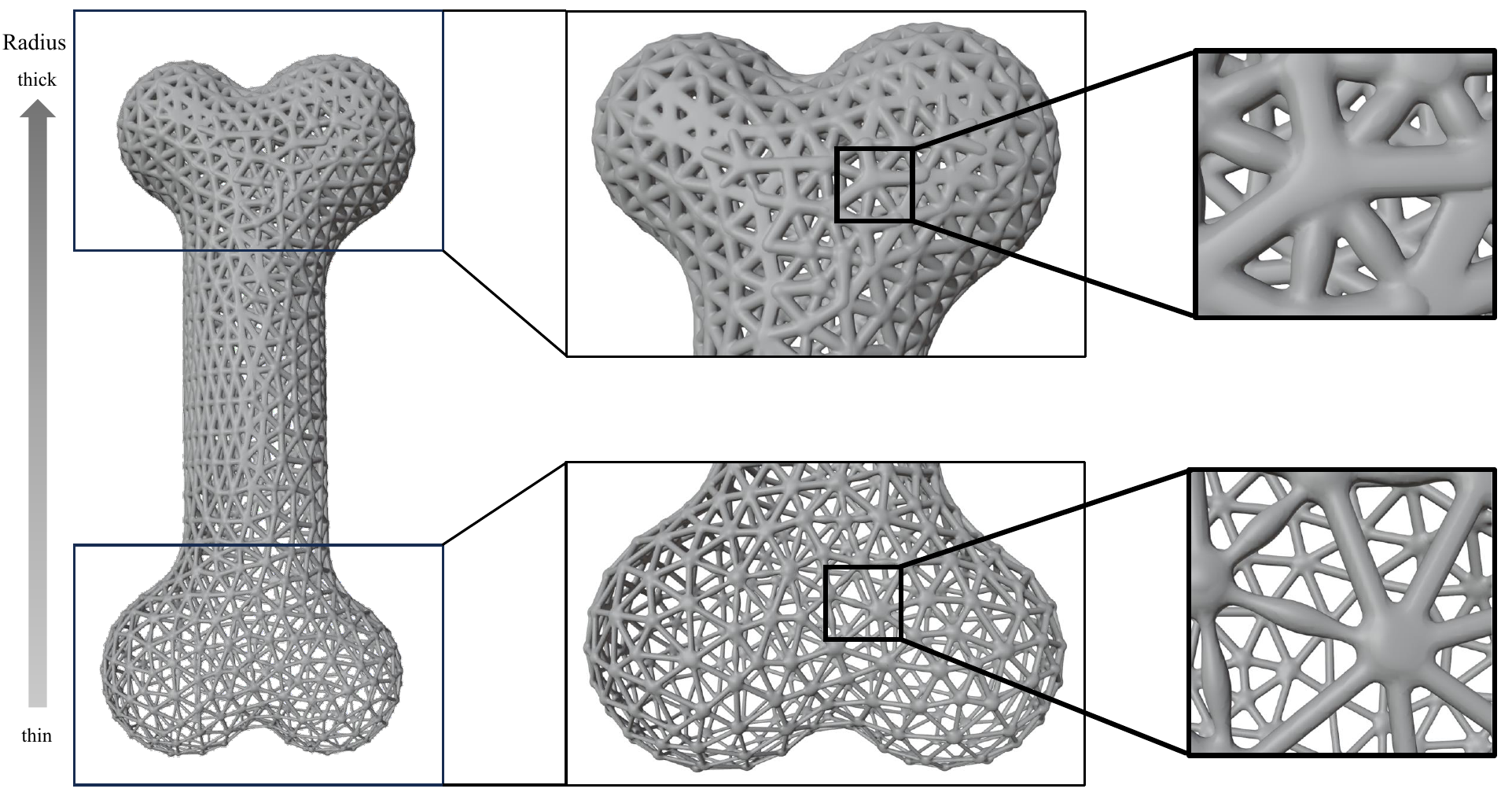}
  \caption{A gradient lattice structure model.}
  \label{fig:gradient}
\end{figure*}

\begin{figure*}[th!]\centering
  \centering
  \includegraphics[width=0.95\textwidth]{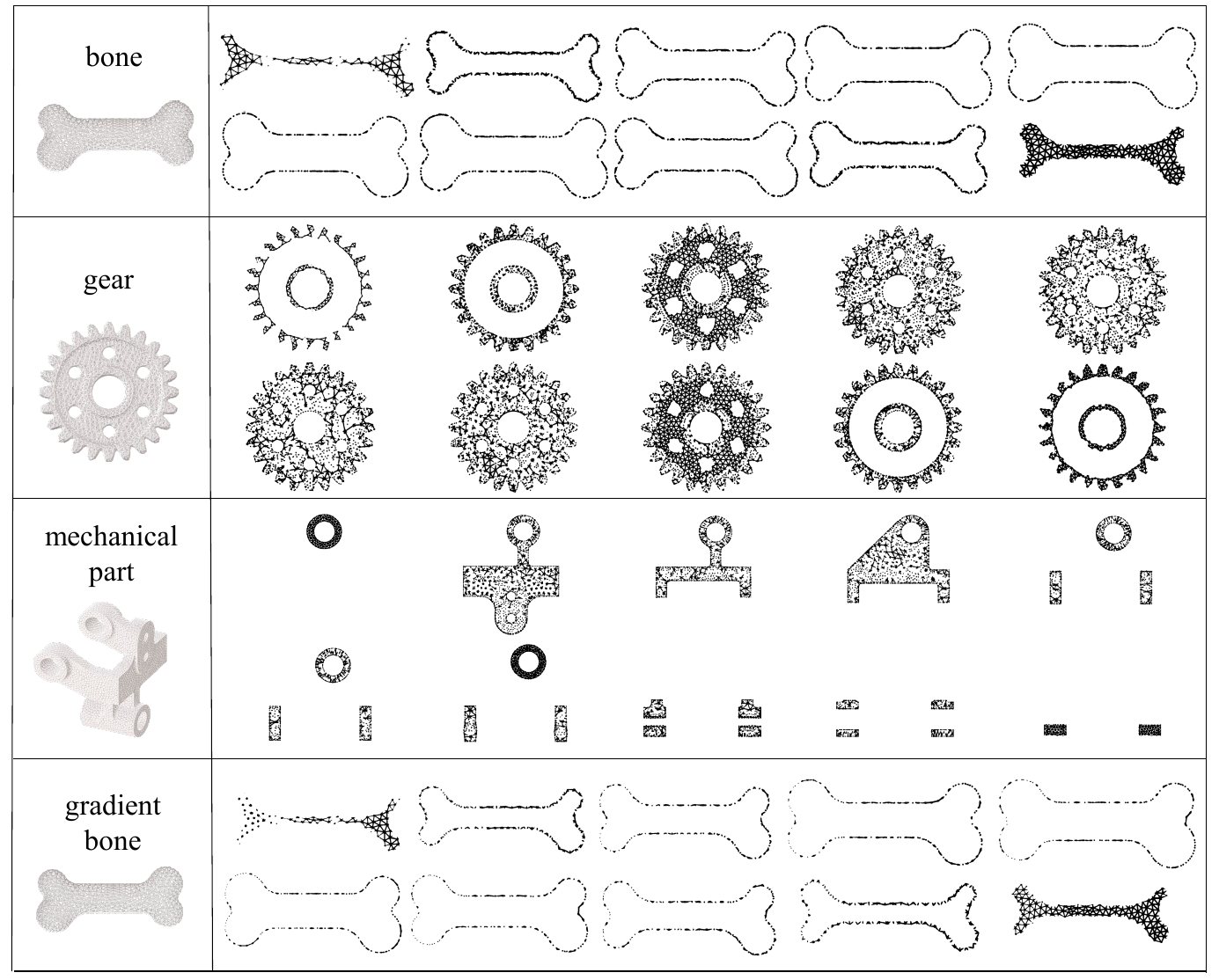}
  \caption{Slicing result of the models in Case Studies 2 and 3.}
  \label{fig:slicing}
\end{figure*}

\begin{table*}[t]
    \centering
    \caption{Slicing time of lattice structure models in Case Study 2 and Case Study 3.}
    \label{tab:slicingtime}
    \footnotesize
    \resizebox{1.0\textwidth}{!}{
    \setlength{\tabcolsep}{6pt} 
    \renewcommand\arraystretch{1.5}
    \begin{tabular}{cc c cccccc}
        \toprule
        \textbf{Model Name} & \textbf{Total Strut Number} & \textbf{Number of Layers} & \textbf{Average Strut Number Per Layer} & \textbf{Resolution} & \textbf{Total Slicing Time (s)} & \textbf{Slicing Time Per Layer (s)}  & \textbf{Model Generation Time (s)} & \textbf{Model Size (MB)}\\ 
        \midrule
        Bone & 2,700 & 211 & 342 & 468*1018 & 12.329 & 0.058 & 17.929 & 234 \\
        \midrule
        Mechanical Part & 37,403 & 416 & 1,394 & 504*593  & 10.108 &  0.024 & 26.289 & 711 \\
        \midrule
        Gear & 22,574 & 209 & 2,904 & 781*781 & 12.446 &  0.060 & 40.386 & 1234 \\
        \midrule
        Gradient Bone & 2,700 & 211 & 342 & 468*1018 & 12.561 & 0.060 & 18.302 & 246 \\
        \bottomrule
    \end{tabular}
}
\end{table*}

To demonstrate the effectiveness of the proposed method, four case studies are provided in this section. Case Study 1 (Fig.~\ref{fig:variable_strut} and Fig.~\ref{fig:double_strut}) demonstrates the diversity of the lattice structure profile shape by providing a set of representative struts with various control parameters, as well as struts constructed by connecting two struts along the axis. Case Study 2 (Fig.~\ref{fig:solid_models}) provides three typical lattice structure models with different control parameters constructed by our method. Case Study 3 (Fig.~\ref{fig:gradient}) shows a bone model with gradient control parameters (e.g., $\mathbf{r}$ in Sec.~\ref{sec:methods}) along the vertical direction, and the zoom-in figure shows the struts with variable radius. Case Study 4 (Fig.~\ref{fig:slicing} and Table.~\ref{tab:slicingtime}) gives the slicing images, slicing time, and generation time of the lattice structures in Case Study 2 and Case Study 3, demonstrating the efficiency and robustness in evaluation. The total slicing time and generation time are tested for 5 runs, and the average outcome is adopted here.

\subsection{Discussion and Limitations}
The above examples and comparisons collectively demonstrate that the proposed method provides a large design space for geometric modeling of lattice structures.
The results shown in Fig.~\ref{fig:variable_strut} and Fig.~\ref{fig:double_strut} indicate that the proposed approach can model lattice structures with various radii within a wide range of control parameters. The results in Fig.~\ref{fig:solid_models} show the practical application in lattice structure design of real-world CAD models. The results in Fig.~\ref{fig:gradient} further demonstrate the parametric control of lattice structures (i.e., the gradient variation) within one model.

From Fig.~\ref{fig:slicing}, we can see that all the slicing images are clear and complete without obvious numerical error or non-manifold situations, demonstrating the robustness of implicit field evaluation. Moreover, the slicing time and the time of generating the models in Case Study 2 and Case Study 3 are provided in Table.~\ref{tab:slicingtime}. All the slicing time per layer is no more than about 0.1 s/layer, which is efficient enough in practical applications. And for model generation, there are two main steps. The first step is slicing every layer, and the second step is applying the Marching Cube algorithm to output the solid model. The total generation time is the sum of these two steps. All the generation times are less than 1 minute, which is acceptable in practical use.

While this proposed method is effective across the presented case studies, the profile shape control could only be conducted indirectly by manipulating the implicit field value, rather than intuitively changing the profile shape. This may lead to insensitivity to the control curves (i.e., the cubic Hermite curve); that is, a relatively large change of the Hermite curve could only bring a little change to the profile shape.

\section{Conclusion}
\label{sec:conclusion}

This paper presented an implicit geometric modeling framework for lattice structures with controllable variable-radius struts.
The proposed method integrates a cubic Hermite curve to control the implicit field, and then further controls the profile shape of struts. Our method increases the design space by using higher-order control curves than traditional linear curves and the state-of-the-art quadratic curves~\cite{2018_GUPTA_quador_hub}.

To further improve computational efficiency, we derived the analytical expression of the convolutional implicit field for a single strut. Since the resulting field is an analytical polynomial function, the evaluation can be conducted efficiently and is suitable for massively parallel computation on GPUs. In addition, by exploiting the superposition property of the adopted convolution surface, more complex profile shapes can be constructed by dividing struts into multiple segments with independent control parameters, which further enlarges the design space of lattice structures.

Three case studies demonstrated that the proposed method can generate a wide variety of strut profile shapes. And the slicing results further verified its efficiency and robustness for practical lattice modeling tasks.

Despite these advantages, the current formulation still has a limitation for building the strut junctions. Since the overall implicit field is constructed by directly summing the field contributions of all struts, regions where multiple struts meet may exhibit an evident bulging effect. This artifact becomes more pronounced when several struts are densely connected, because the accumulated field values can locally enlarge the resulting isosurface. Although such a representation remains robust and easy to evaluate, the bulge may reduce geometric fidelity at lattice nodes in applications like computer animation. A possible direction for future work is to evaluate the implicit field of each strut separately and then combine these fields using blending operators, rather than simple summation. Such a strategy may help suppress excessive field accumulation at junctions while preserving smoothness between adjacent struts, and therefore deserves further investigation in future research.

\section*{Acknowledgements}

This work has been funded by the ``Pioneer" and ``Leading Goose" R\&D Program of Zhejiang Province (No. 2024C01103), the National Natural Science Foundation of China (No. 62102355), and the Fundamental Research Funds for the Zhejiang Provincial Universities (No. K20241957, K20250142).

\bibliographystyle{elsarticle-num}
\bibliography{mybibfile}


\end{document}